\documentclass[twocolumn,showkeywords,preprintnumbers,amsmath,amssymb]{revtex4}
\usepackage{graphicx}% Include figure files
\usepackage{dcolumn}% Align table columns on decimal point
\usepackage{bm}% bold math

\begin{document}

\title{Physical Version of Singularity Resolution in the Observable Universe}% Force line breaks with \\

\author{Daegene Song}

\affiliation{%
Department of Management Information Systems, Chungbuk National University, Cheongju, Chungbuk 28644, Korea
}%

\date{\today}% It is always \today, today,
             %  but any date may be explicitly specified

\begin{abstract}

Based on the equivalence of the two different types of measurement protocols and the 
asymmetry between the Schr\"odinger and Heisenberg pictures, 
it has been  previously proposed that negative sea fills the universe as a 
nondeterministic computation - a time-reversal 
process of the irreversible computations presented since the big bang. 
The goal of this paper is to extend the proposed subjective universe model, i.e., the universe 
as a quantum measurement: 
Motivated by the relationship between quantum theory and classical probability theory 
with continuity, it is argued that the frame of reference of the observer may be identified with classical probability theory 
where its choice, along with big bang singularity, 
should correspond to the quantum observable.  
That is, the physical version of singularity resolution corresponds to the case,  
where big bang singularity is equivalent to the continuity of the negative sea, or aether, 
filling the universe as a frame of reference of the observer. 
Moreover, based on the holographic principle, we identify 
the choice of the observer with the 
degrees of freedom proportional to the Planck area on the horizon. 
We also discuss that the continuity or infinity present 
in every formal language of choice acceptable in nondeterministic computation 
may be associated with the universal grammar proposed by Chomsky in linguistics.

\end{abstract}

\maketitle

\section{introduction}

A central issue in physics has been the 
apparent discrepancy between the classical and quantum worlds. 
Indeed, subatomic particles, such as photons or electrons, 
exhibit peculiar behavior, such as a single photon moving through 
two different paths at the same time, which is unseen in the classical world.  
Therefore, why do such odd phenomena generally occur  
on small scales, i.e., at the microscopic level rather than the macroscopic level such as  
in buildings, stars, etc.? 
There have been a number of suggestions associated with  
this dilemma between the quantum and classical worlds including 
the example of decoherence \cite{zurek1,zurek2}.
However, no conclusive consensus has been reached among researchers to date.

In \cite{hardy}, it was shown that one can derive quantum theory 
from a set of simple axioms.  In particular, 
the removal of one axiom - continuity - is equivalent to the classical probability theory.
In this paper, we will argue that, when we take classical probability 
as the choice of the observer, with continuity imposed 
on the classical choice, it should correspond to the quantum observable. 
In particular, we will discuss how classical probability theory, as 
the observer's choice, can be identified  with the 
degrees of freedom lying on the horizon, and that the continuity 
area corresponds to the negative sea or, as suggested 
in \cite{song8}, the aether, which fills the universe. 
 This is rather surprising because, for many years, people have 
often considered the classical as an approximation of quantum theory.  
However, quantum theory does not exclude the classical world.  In fact, classical spacetime is an 
integral part of standard quantum theory because it contains 
not only unitary transformation but also measurement, 
where the latter is completed in classical spacetime.

\begin{figure}
\begin{center}
{\includegraphics[scale=.45]{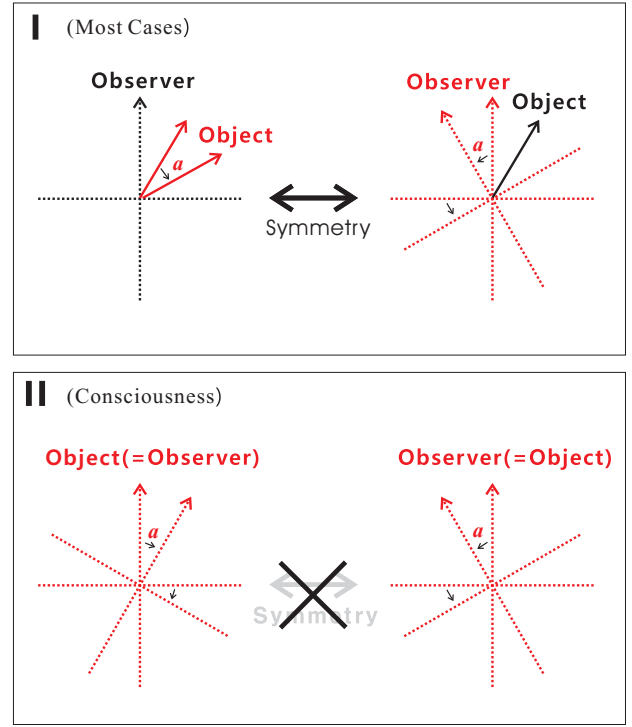}}

\end{center}
\caption{ [i] Most physical phenomena obey the symmetry 
between the Schr\"odinger and the Heisenberg pictures.  [ii] In the case of 
consciousness, this symmetry breaks down.  }
\label{consciousness}\end{figure}

In sect. 2, we review the previously proposed equivalence 
between the two-system and single-system protocols with negative sea. 
In sect. 3, we will argue that the choice of the observer in 
measuring the observable universe may be described by the classical probability 
theory with continuity, which is equivalent  
to the quantum observable.  We will then conclude with brief remarks.

\section{Asymmetry and Quantum Measurement}
In \cite{song1,song3}, the subjective nature of existence 
was motivated by the contradiction that appeared in the 
self-observation of consciousness.  The precision of advancements 
in physics, which previously attempted to create 
an objective rule for physical systems, finally led to 
the description between the observing party and the object 
shown in quantum theory at the beginning of the 20th century. 
This advancement, which exhibited subjectivity, was not easily 
accepted by many researchers at the time \cite{EPR}.

\begin{figure}
\begin{center}
{\includegraphics[scale=.65]{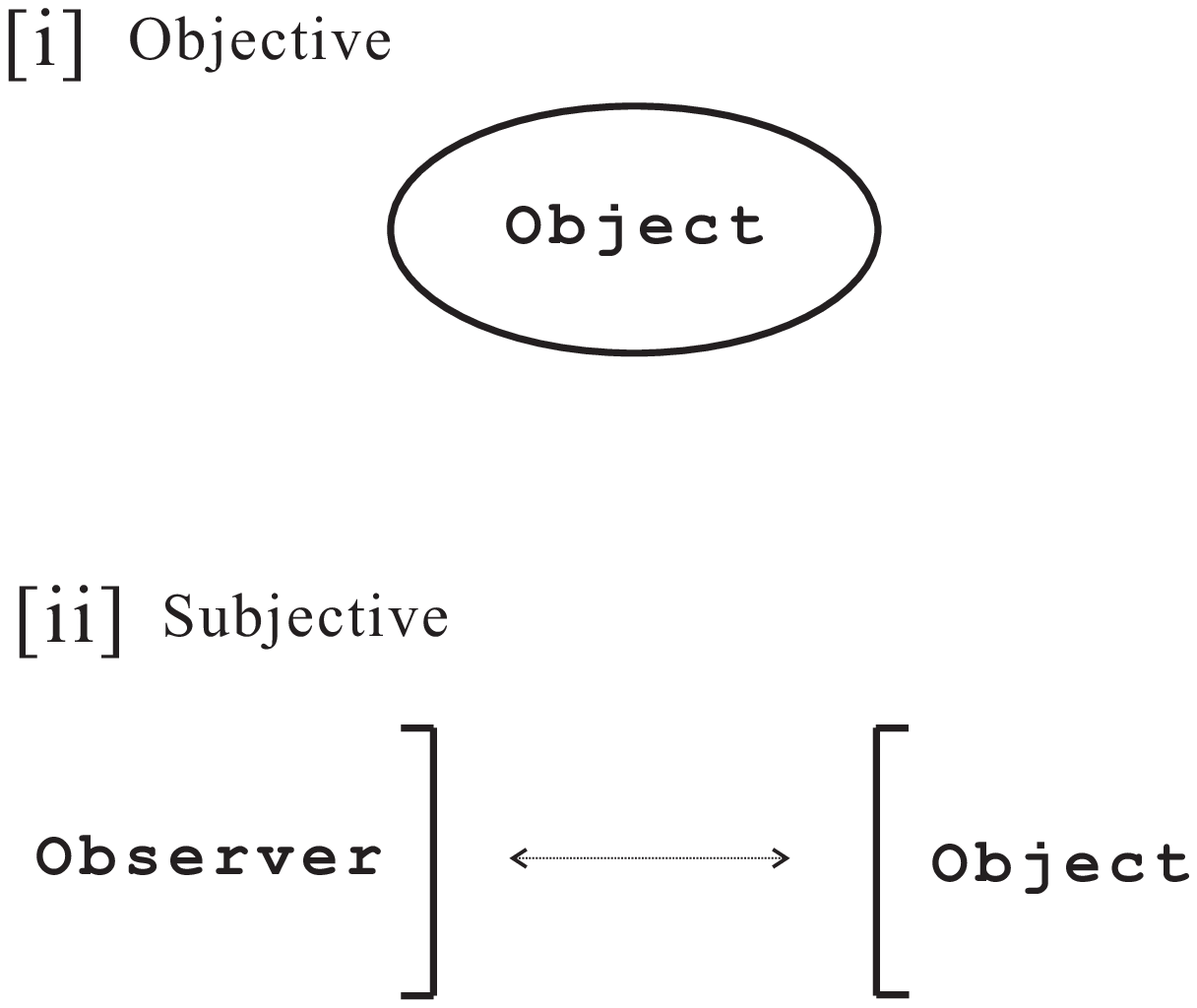}}

\end{center}
\caption{ Paradigm shift: Previously, finding the objective pattern of a given object was pursued [i].  However, 
with quantum theory, the relation between the observer and the object is studied instead [ii]. 
  }
\label{Subj}\end{figure}

In particular, it was argued \cite{song2} that one may consider the 
observable as the frame of reference of the observer, when observing 
the given quantum system.  This postulate 
leads to an asymmetry between the Schr\"odinger and Heisenberg pictures, i.e., 
active and passive transformations, respectively, when 
the very object being observed is the frame of reference itself - a phenomenon 
that only occurs in consciousness (Fig. \ref{consciousness}). 
 It was then argued that, to successfully keep this inconsistency from 
 occurring in the case of consciousness, 
the basic assumption of treating the observer and the object 
separately, when considering the measurement to be the 
relative difference between the two, should stop. 
Instead, the two entities are not separable, and their existence 
should be subjective.

Conversely, another highly 
debated subject is in regard to free will. 
This is due to the deterministic worldview, which was held 
until the development of quantum theory and 
often regarded as having no place for the nondeterministic aspect of free will.  
However, with randomness as an essential 
ingredient of the theory, many suspected quantum theory may 
open the possibility of the existence of free will \cite{FW1,FW2,FW3,FW4}.  
Nevertheless, free will has not only randomness, 
but two seemingly contradictory aspects instead, i.e.,  
\begin{enumerate}
\item from a subjective perspective, the observer is able 
to freely choose; 
\item to the outside, the choice ought to be 
unpredictable and random. 
\end{enumerate}
That is,  with all the initial conditions known about the observer, the choice should 
be random to the outside; however, from the subjective aspect, the observer is free to choose. 
The next section will review the theory that the subtlety involving free will 
may be physically realized using a nondeterministic computation.

Motivated by the black hole information problem \cite{hawking} 
and two different measurement protocols in quantum theory, 
it was argued \cite{song5} that 
 the process of black hole radiation should be 
considered as a quantum measurement. In particular, it was shown that 
the observer's free will outside the black hole results from 
the choices made inside the horizon, with 
the memory state, as follows:  
\begin{equation}
\mathcal{O}_{\theta}^{out} = Q_{\theta}^{in}
\label{TheEquation}\end{equation}
That is, the observer's choice is hidden behind 
the horizon yet fills the vacuum outside  the horizon with 
negative information, which may be considered the consciousness of the observer. 
It is interesting to note that free will, when used with black 
hole entropy, indeed has the dual aspects discussed above.

\begin{figure}
\begin{center}
{\includegraphics[scale=.6]{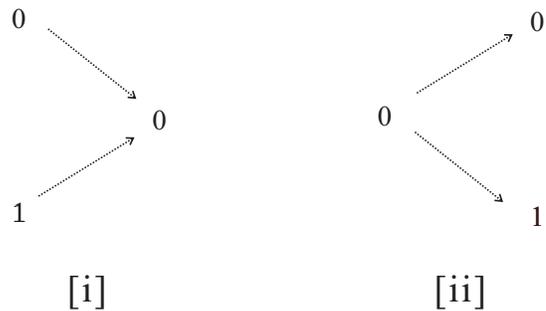}}

\end{center}
\caption{ [i] Irreversible computation: Given the output, 
it is impossible to determine the trail back to the input.  Landauer has shown \cite{landauer} 
that this process necessarily dissipates energy.  
[ii] Nondeterministic computation (or free will): The time-reversal process of the irreversible computation in [i]. }
\label{nondeterministic}\end{figure}

Notably, this picture is also consistent with the subjective 
approach in quantum theory, particularly the Copenhagen interpretation. 
While the traditional approach in physics has been to 
find an objective pattern of a given physical system, the 
subjective approach attempts to provide a relationship 
between the observing party and the object, i.e, with the observable 
and the state vector, respectively (Fig. \ref{Subj}).  The above equivalence 
of quantum measurement protocols and black hole evaporation indeed provides an 
explanation of how the special status of the 
observable in the subjective approach arises, i.e., by considering the 
objective observing party, or the apparatus, 
as traveling backward in time or negative sea filling the vacuum.

\section{Classical \& Quantum}  
\begin{figure}
\begin{center}
{\includegraphics[scale=.6]{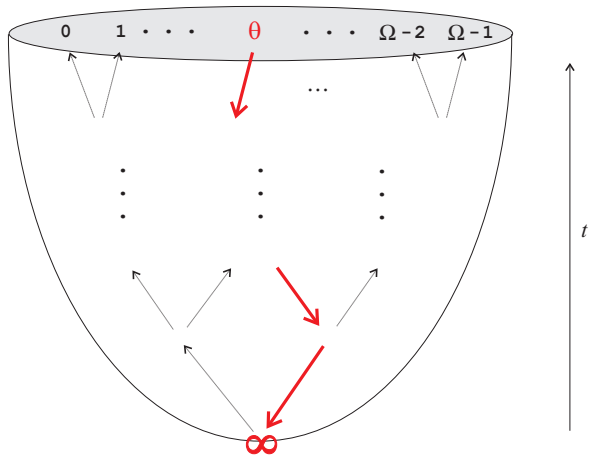}}

\end{center}
\caption{The universe as quantum measurement: By considering the universe as a computational process, 
the subjective model suggests that 
the negative sea, which corresponds to the time-reversal process 
of irreversible computation, fills the universe,  
 where $\Omega$ is the 
number of equally accessible microstates of the observable universe and $\theta$,  
$0\leq \theta \leq \Omega -1$, is the measurement choice made by the observer.   }
\label{DiracSea}\end{figure}

The equivalence of two different 
measurement protocols in (\ref{TheEquation}) has been extended 
to the cosmological model \cite{song7}.  In \cite{lloyd}, the universe was modeled 
as a computation process, and the maximum number of 
possible irreversible computations 
since the big bang has been estimated based on the Margolus-Levitin theorem \cite{margolus},  
which suggests the minimum time required to perform elementary gates equals $\frac{\pi\hbar}{2E}$. 
Based on this computational model of the observable 
universe, it was elaborated \cite{song7} how the observer's choice, or free will,  
may play an essential role in building the specific model of the universe by 
using a nondeterministic computation (Fig. \ref{nondeterministic}). 
That is, by viewing the universe as a computational process, it was argued that 
the entropy of the observable universe corresponds 
to the number of computations of nondeterministic computation, which is a reverse process 
of irreversible computation, such that it fills the 
vacuum as a Dirac-type negative sea, as shown in Fig. \ref{DiracSea}.  
\begin{flushleft}
{\it{The observer's choice is a nondeterministic computation 
that travels backward in time all the way to big bang singularity.  }}
\end{flushleft}

The nondeterministic computation 
chooses the acceptable path of computational processing, 
which is different from probabilistic computation. 
This should correspond to the observer's subjective 
experience of making choices freely, i.e., 
rather than randomly, as in a probabilistic computation, which 
fits the first criteria among the dual aspects of free will discussed earlier (also see \cite{song6}).

As a result, the entropy of the observable universe 
corresponds to a logarithm of possible choices that the observer is able to choose.  
Indeed, it was argued, that for any $\Omega$ equally accessible 
microstates of the universe:     
\begin{flushleft}
{\it{The observer's choice corresponds to the reality of the universe.  }}
\end{flushleft}
Therefore, 
the subjective universe model, i.e., the universe as a quantum measurement, 
which is proposed in \cite{song7},  
suggests that the observer's freely chosen will is the actual 
existence with the dual aspect of free will as well.

\begin{figure}
\begin{center}
{\includegraphics[scale=.6]{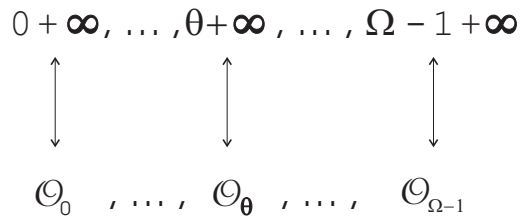}}

\end{center}
\caption{The choice of the observer in classical spacetime among the equally probable 
$\Omega$ can be made equivalent to the choice of the quantum observable.  
In particular, the continuity, or $+\infty$ present in every 
choice of the formal language $0\leq \theta \leq \Omega -1$, should correspond to 
the universal grammar proposed by Chomsky.   }
\label{Corres}\end{figure}

In \cite{hardy}, it was shown that quantum theory can be derived from a set of axioms.  
In particular, it was argued that, with the same set of axioms - 
except continuity - they yield classical probability theory. 
Therefore, if we use the notation $+\infty$ to represent the continuity axiom \footnote{For example, see 
the discussion in \cite{galvao} for the connection between continuity of a qubit and the infinite number of 
classical bits.}, we may write: 
\begin{equation}
{\rm{Class. \, Prob.}} + \infty \Longleftrightarrow  {\rm{Quantum}}
\label{hardy}\end{equation}
When one performs a measurement of a quantum state, 
an observable is used, where both the state 
and the observable are associated with complex vector space. 
However, the observer does not have direct access to this vector space but rather only 
to a classical frame of reference, which is defined by classical 
spacetime.  
Therefore, when we refer to the classical frame of reference, 
we wish to identify it as the frame of reference in spacetime
that has a corresponding quantum observable.

Returning to the universe model in Fig. \ref{DiracSea}, 
let us define the observer's equally probable choice $\theta$ to be 
the choice of the classical frame of reference, 
where $0\leq \theta \leq \Omega -1$, and $\Omega$ is the number of equally accessible microstates of the 
universe. 
With this identification and following (\ref{hardy}), the classical choice of $\theta$ with continuity 
may be considered as equivalent to the degenerate quantum observable $\mathcal{O}_{\theta}$.  
Therefore, as shown in Fig. \ref{Corres}, we will use the notation 
$\theta + \infty$ as the equivalent of the observable $\mathcal{O}_{\theta}$, i.e., 
\begin{equation}
\theta + \infty \Longleftrightarrow \mathcal{O}_{\theta}
\label{theta}\end{equation}

\begin{figure}
\begin{center}
{\includegraphics[scale=.65]{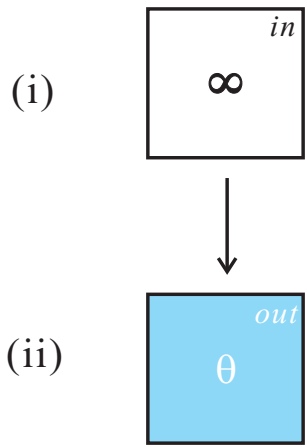}}

\end{center}
\caption{The physical version of singularity resolution: Big bang singularity, with the choice 
of the observer, can be considered as equivalent to the 
observer's classical choice $\theta$ with continuity, or $\mathcal{O}_{\theta}$, and as  
negative sea filling the observable universe.   }
\label{resolution}\end{figure}

In particular, the classical choice of $\theta$ with 
equal probability $\frac{1}{\Omega}$ corresponds to the horizon degeneracy conjectured 
from the holographic principle \cite{susskind}.  
The Bekenstein-Hawking entropy \cite{bekenstein,hawking1} corresponds to: 
\begin{equation}
S_{\rm{BH}} = \frac{kA}{4l_p^2}
\label{BH}\end{equation}
where $k$ is the Boltzmann constant, $A$ the area of 
the horizon, and $l_p$ is the Planck length.  Conversely, 
Boltzmann's entropy law yields a logarithm of the number of possible configurations, 
\begin{equation}
S_B = k \ln \Omega
\label{boltzmann}\end{equation}
The holographic principle states that the degrees of 
freedom inside are encoded on the horizon surface.  That is, 
the number of possible internal degeneracies corresponds to 
\begin{equation}
\Omega = e^{\frac{A}{4l_p^2}}
\label{degree}\end{equation}
or $\sim 1$ bit per Planck area is encoded on the horizon.  
By following the argument of the holographic principle and our suggestion 
of classical probability with continuity, we suggest that: 
   
\begin{flushleft}
{\it{The choice $\theta$ of the observer with continuity, or $+\infty$,  
has the classical degrees of freedom residing on the horizon. }}
\end{flushleft}

It should be noted that the above claims that the degrees of 
freedom on the horizon, as in the holographic principle,  
correspond to the classical domain. 
Ever since the discovery of black hole radiation \cite{hawking1}, 
the statistical nature of entropy has been debated by scholars;   
however, the above suggestion of horizon entropy, which corresponds 
to the classical configuration, is consistent with the 
statistical calculation of the entropy, as shown in \cite{gibbons}.

A primary area of research, in an attempt to understand the mental 
process, has been the study of human language. 
In particular, linguist Chomsky has claimed \cite{chomsky1,chomsky2} 
that there is a universal structure 
in all languages, which is neither learned nor acquired by experience. 
This concept came to be known as universal grammar, and its innateness remains 
controversial and debated \cite{UG1,UG2}. 
If we consider the choice of the observer, which may be represented 
in binary bits, a formal language (\ref{theta}) implies that 
every language, i.e., $0\leq \theta \leq \Omega -1$, contains 
the continuous part, or $+\infty$. This property is in every language 
$\theta$, and it is consistent with the proposal of universal grammar.  
That is, while formal language is written as a finite combination of classical bits, 
it always contains the continuous or infinite (Fig. \ref{Corres}) 
conscious aspect dominated by the quantum 
theory of negative sea.

\section{Remarks}
In this paper, we have provided a more specific model of the subjective universe model proposed 
in \cite{song2,song3}, i.e., the observer and the object are not separable.  It was 
discussed that the choice of the observer in classical spacetime 
with continuity fills up the universe as negative sea.  Moreover, the classical 
choice of the observer has degrees of freedom on the horizon, 
which are proportional to the Planck area and the 
continuous negative sea, or the aether, serve as a conscious 
frame of reference of the observer; this  
may be considered as a resolution of big bang singularity, as shown in Fig. \ref{resolution}.  
It was also discussed that 
the continuity part, or $+\infty$, which is present in every formal language,  
as shown in Fig. \ref{Corres}, should correspond to the universal grammar proposed in 
linguistics.

Moreover, the above argument suggests that discrete spacetime at the 
Planck level follows classical probability rules, i.e., as the frame of reference  
of the observer, yet each classical degree of freedom is 
associated with continuity, which leads to 
quantum theory. 
This picture is in fact consistent with quantum theory,  
which has two components: the 
unitary transformation, which occurs in complex Hilbert space, 
and the measurement sector, which occurs in classical spacetime.

{\it{Acknowledgments}}:  This work was supported by Research Funding 
program of Chungbuk National University.

%%%%%%%%%%%%%%%%%%%%%%%%%%%%%%%%%%%%%%%%%%%%%%%%%%%%%%%%%%%%%%%%%%%%%%%%
%%%%%%%%%%%%%%%%%%%%%%%%%%%%%%%%%%%%%%%%%%%%%%%%%%%%%%%%%%%%%%%%%%%%%%%%
%%%%%%%%%%%%%%%%%%%%%%%%%%%%%%%%%%%%%%%%%%%%%%%%%%%%%%%%%%%%%%%%%%%%%%%%
%%%%%%%%%%%%%%%%%%%%%%%%%%%%%%%%%%%%%%%%%%%%%%%%%%%%%%%%%%%%%%%%%%%%%%%%
%%%%%%%%%%%%%%%%%%%%%%%%%%%%%%%%%%%%%%%%%%%%%%%%%%%%%%%%%%%%%%%%%%%%%%%%

\end{document}